# Synergistic effect of graphite nanoplatelets and carbon black in multifunctional EPDM nanocomposites

L. Valentini,[*] S. Bittolo Bon

*Dipartimento di Ingegneria Civile e Ambientale, Università di Perugia, UdR INSTM, Strada di Pentima 4, 05100 Terni - Italy. Tel: +39 0744 492924; E-mail: luca.valentini@unipg.it*

M. A. Lopez-Manchado, R. Verdejo

*Instituto de Ciencia y Tecnología de Polímeros, ICTP-CSIC, Juan de la Cierva, 3 28006 Madrid - Spain*

A. Bolognini,

*Dipartimento di Fisica e Laboratorio SERMS, Università degli Studi di Perugia, Strada di Pentima 4, 05100 Terni - Italy*

A. Alvino, S. Borsini

*SERMS srl, Strada di Pentima 4, 05100 Terni – Italy*

A. Berardo

*Laboratory of Bio-Inspired and oGraphene Nanomechanics, Department of Civil, Environmental and Mechanical Engineering, University of Trento, Trento - Italy*

N. M. Pugno

*Laboratory of Bio-Inspired and oGraphene Nanomechanics, Department of Civil, Environmental and Mechanical Engineering, University of Trento, Trento - Italy*

*Center for Materials and Microsystems, Fondazione Bruno Kessler, Trento - Italy*

*School of Engineering and Materials Science, Queen Mary Univerisity of London, Mile End Road, London - United Kingdom*

**Abstract**

Ethylene–propylene–diene terpolymer rubber (EPDM)-based nanocomposites containing carbon black (CB), graphite nanoplatelets (GNPs), and mixtures of the two fillers were prepared. The influence of the relative amounts of the two fillers on the dynamic and static friction coefficients was examined. The static analysis of the coefficient of friction suggests that the partial substitution GNPs into the EPDM/CB blend did not produce a significant variation of the surface grip. The sample comprising EPDM/CB composite and an effective amount of GNPs dispersed in the matrix provides an increase of the thermal conductivity, damping and mechanical properties of the nanocomposites. The morphological observations revealed that the replacement of CB with GNPs reduces the CB aggregation and, hence, improving the percolation of the hybrid fillers and the interface resistance of the composite. The development of thermally conducting elastomeric nanocomposites could envisage their utilization in the processing of rubber blends satisfying the increasing demand to reduce both the duration of the vulcanization process and thus the cost of the vulcanized rubbers.

**Introduction**

Rubber is commonly considered the workhorse of the industrial and automotive products because of its good mechanical properties and its relatively low cost; finished products are found in the market place as compression molded products. The physical and chemical resistance properties of rubber materials are determined by the addition of carbon black (CB) that historically has been utilized to reinforce rubber matrices [1,2].

However, since the main factors that affect the composite properties are the particle size and the mode of interactions with the matrix materials, single filler is not generally able to meet the structural and functional requirements of rubber composites [3-7]. For example, it was found that the partial replacement of CB with carbon nanotubes leads to a much lower percolation threshold than that of the composite obtained with single filler and to a synergetic effect on the composite mechanical properties [8,9].

It is widely acknowledged that, in some cases, a combination of two or more carbon fillers could improve the electrical performance of the composite due to the synergistic effect [10-12]. ]. Ma et al. [13] reported that the addition of carbon nanotubes into the composites filled with CB could remarkably enhance the electric conductivity of the matrix, and result in a low percolation threshold of 0.4 wt.%. Recently, Yang et al. [14] studied the effects of substituting CB with graphene oxide/CB and reduced graphene oxide/CB hybrid fillers on the structure and properties of natural rubber composites.

In addition to graphene oxide, multi-layer graphene platelets also exhibit unique and useful behaviors. Multi-layer graphene, herein referred to as graphite nanoplatelets (GNPs) contains essentially no oxygen (<1% by weight of oxygen). Graphite nanoplatelets (GNPs) are obtained from expanded graphite via rapid evaporation of the intercalated compounds at elevated temperatures. The extent of thermal expansion (and therefore the platelet thickness) is dependent on type of graphite used, beyond that on intercalation procedure [15,16]. With such a method, the graphite nanoplatelets so obtained typically consist of hundreds of stacked graphene layers and average between 30 and 100 nm in thickness.

Thus, the key to utilizing graphite as nanoreinforcement relies on the ability to exfoliate graphite compounds. Since the cost of natural crystalline graphite is quite cheap, around 1.5-1.6 $ lb$^{-1}$, the cost of exfoliated graphite is expected to be 5 $ lb$^{-1}$ or less. This is significantly less expensive than carbon nanotubes (~7500 $ lb$^{-1}$) or carbon fibers (40-50 $ lb$^{-1}$), yet the properties of crystalline graphite flakes are comparable to those of nanotube and carbon fibers [17] .

The 2D nanoscale dimension of GNPs is a huge benefit in relation to the large conventional 3D fillers [18]. Those graphitic inclusions are characterized by far better shape factor, larger contact surface and higher mechanical strength. Whereas, the strong intrinsic van der Waals attraction between the sheets and the high surface area makes the GNPs easily aggregate and difficult to disperse in the matrix, the synergy among the hybrid fillers comprising of graphite intercalation compounds, mainly GNPs, and CB could lead to the development of graphite-based elastomer composites exhibiting exceptional mechanical and thermal properties.

It is known that rubbers or elastomers generally have a low thermal conductivity. Consequently, when such materials are used as packaging for electronic circuit, they store the generated heat that in turn raises the temperature of the device itself, thereby promoting heat deterioration of the electronic component. To achieve this goal, the heat conduction capability of a rubber may be improved by compounding a rubber with a filler having a heat conductivity higher than that of the rubber.

High filler loadings (>30 vol.%) were typically necessary to develop functional EPDM elastomers with appropriate level of thermal conductivity [19]. Indeed, the processing requirements, such as possibility to be extruded and injection molded, often limit the amount of fillers in the formulation and, consequently, the thermal conductivity performance. Moreover, high filler loading dramatically alters the polymer viscosity and density. For these reasons, obtaining rubber composites having thermal conductivities and usual mechanical properties is very challenging at present [20]. Furthermore, traditional metallic materials with the highest thermal conductivity are too heavy and subjected to corrosion.

In this study, the development of hybrid fillers system consisting of GNPS and CB was reported. The effects of substituting GNPs for CB on the thermal, damping and mechanical properties of rubber/CB composites was studied and rationalized in terms of the morphological analysis.

**Experimental details**

Ethylene-propylene diene terpolymer rubber (EPDM) was kindly supplied by Exxon Mobil Chemical under the trade name Vistalon 7500 (ethylene content: 56.0 wt.% and 5-ethylidene-2-norbornene (ENB) content: 5.7 wt.%). Carbon black was kindly supplied by Cabot, S.A. under the trade name Vulcan 3-N330 (diameter 225 nm with a surface area of 77 $m^2$/g) and a paraffinic oil kindly supplied by Nynas, Nyflex 820 was used as plasticizer. Graphite nanoplatelets, an intermediate grade between graphene and graphite, which can neither be considered pure graphene nor graphite were purchased from Cheaptubes.

Rubber compounds were prepared in an open two-roll mill at room temperature. The rotors operated at a speed ratio of 1:1.4. The vulcanization ingredients were sequentially added to the elastomer before to the incorporation of the filler and sulphur. The recipes of the compounds are described in Table 1. Vulcanizing conditions (temperature and time) were previously determined by a Monsanto Moving Die Rheometer MDR 2000E. Rubber compounds were then vulcanized at 160 ºC in a thermofluid heated press. The vulcanization time of the samples corresponds to the optimum cure time $t_{90}$ derived from the curing curves of the MDR 2000E. Specimens were mechanically cut out from the vulcanized plaques. Field emission scanning microscopy (FESEM) was used to investigate the cross section of the samples.

Tensile stress–strain properties were measured according to ISO 37–1977 specifications, on an Instron dynamometer (Model 4301), at 25 ºC at a crosshead speed of 500 mm*min$^{-1}$. At least five specimens of each sample were tested.

A ball-on-disk tribometer was used to determine the dynamic friction of coefficient of the composites. The samples were cut in order to have a squared base with different measures, from 8x8 mm$^2$ to 15x15 mm$^2$ (average values), depending on the given materials. They were fixed in the tribometer and the antagonist material we chose was steel (100Cr6), a sphere of 6 mm diameter in order to have a single contact point between the rubber and the steel. No lubricants were used. The sliding velocity was set at 0.01 m/s and the normal load varied from 0.05 N (softer samples) to 0.1 N (harder samples). For each sample from three to five measurements were realized.

The method used to measure the static friction coefficient is based on the Coulomb theory of friction. Each sample was positioned on a plate and fixed on it. After, a weight is put on the sample. The plate was then tilted until the stable configuration was overwhelmed and the weight slides on the rubber surface. The final configuration is tilted by a certain angle with respect to the initial position of the plate and corresponds to the transition from a stable state (static equilibrium) to an unstable one (incipient movement). The tangent of that angle corresponds to the ratio between the tangential force and the normal applied load (the weight). Five measures per sample were performed.

The damping properties were tested through a vibration generated via a pneumatic percussion system hitting a metallic plate. The impact area is a metallic plate where the sample to be tested has been fastened to. The sample was hit by a percussion which excites the vibration. A shock accelerometer positioned in the back plate is thus excited and the response is recorded and digitalized via high performance data acquisition system. The impact velocity was set to 8m/s resulting in an impact energy of 58J. Three tests were repeated on each sample; the experimental error was estimated below 1%.

Thermal conductivity measurements follow the "two thermometer-one heater" method using a custom built stage. Two PT100 thermocouples, contacted to the surfaces of a 13*40 mm$^2$ rectangular shape and 14 mm thick sample, monitor the temperature of two polished oxygen-free sample sides. A 3,4 Ohm resistor heats the top plate (13*40 mm$^2$ surface, 14 mm thick) to a temperature $T_{Hot}$ . Heat flows from the top plate, through the sample, and into the bottom plate which is thermally grounded to $T_{Cold}$ (i. e. 20°C) by the cold plate. Thermally conducting grease was used to enhance the thermal contact to the bottom of the sample. A Mylar cap around the cold plate fixed at $T_{Cold}$ and a high vacuum $10^{-5}$ Torr reduce thermal losses due to radiation and convection, respectively.

**Results and discussion**

The tensile properties given in terms of the modulus at different strains (50%, 100% and 300%) and the maximum strength are reported in Figures 1 and 2. It is known that carbon blacks or silica when added to elastomers create a modulus that increases with strain. This non-linearity protects rubber from damage during large deformations [21]. Pristine GNPs provide enhanced non-linear strain hardening to elastomers. The interface is similar to that of carbon black, but the flexibility of the GNPs enables deformation at low strains and hardening at higher deformations. As expected, the addition of the fillers to the EPDM matrix gives rise to an increase of the stiffness of the material which is reflected in an improvement of the modulus at different strains (Figure 1). In particular, the sample EPDM-6 (i. e. 2 wt.% of GNPs and 24wt.% of CB) showed a higher increment of the maximum strength (Figure 2).

The dynamic friction coefficients of the samples were estimated accordingly to the Herzian analysis for a smooth sphere in contact with a smooth flat surface, where the radius of contact circle expressed as $a=[3LR/4E]^{1/3}$, where L is the applied load, R is the sphere radius and E is the elastic modulus of the softer material (i. e. rubber). In the present case the only parameter varied was the load, thus accordingly to the mechanical properties, it was decreased for the softer composite samples containing a GNP/CB ratio of 2/0, 5/0 and 10/0, respectively. The final values are shown in Figure 3. For composites with a GNP/CB ratio of 5/0, 10/0 and 2/24 values major than 1 were obtained and in literature for particular combinations of rubbers similar results were found (i. e. rubber-steel contact) [22-26]. It was also reported that the dynamic friction coefficient depends on the sliding velocity, it increases if the velocity increases, but become almost stable for velocities from 0.01 m/s and more [23-26].

The static coefficient of friction of the samples was estimated by putting a weight made of steel (0.7 g) on the rubber samples and tilting the plate, until the incipient sliding was reached. The dynamic and static coefficients of friction are not comparable due to the different type of steel used as counterpart as well as the different type of setup adopted for dynamic and static tests. The addition of GNPs to the EPDM/CB blend reduces the static coefficient of friction while the partial substitution of CB with GNPs did not affect the grip of the sample surfaces.

Figure 4 reports the peak acceleration measured in the impact excitation test. The damping of the sample can be qualitatively estimated by the peak acceleration. It is evident how the damping performances depend on the synergistic combination of GNPs and CB. In particular, the samples without CB show a scarce damping. The combination of GNPs and CB in the sample with 2 wt.% of GNPs and 24 wt.% of CB showed the best damping with a lower variation of the acceleration peak after the impact. The further increase of the GNPs content to 5 % in the 48 wt.% CB filled matrix deteriorates the damping properties. The obtained results can be explained with the increase of the modulus at different strain along with the reduction of elongation when the GNPs were added.

Figure 5a shows the experimental set up for the thermal conductivity measurements. It is commonly believed that the thermal conductivity of the filled conductive polymer derives from the formation of a conductive network by the fillers in the matrix, and the increase of conductive paths facilitates the improvement of the composite thermal conductivity [27]. As for the CB filler alone, the conductive network is formed due to the contact of the CB agglomerates with each other. For the sample EPDM-6 (i. e. 2 wt.% GNPs and 24 wt.% CB) when GNPs are added into the CB, GNP particles act as a "spacer" and can decrease the agglomeration of the CB, which is favourable to the formation of more conductive paths (Figures 5b and 6a). FESEM analysis reported in Figure 6a supports this hypothesis showing small CB agglomerates attached on the surface and edge of the GNPs; small and well dispersed CB agglomerates for the 2 wt.% GNPs/24wt.% CB ratio can also effectively link the narrow gaps between the GNPs resulting in the formation of additional conductive paths and increasing the interface resistance in the hybrid composite. On the other hand, increasing the GNPs content into the sample with the highest CB concentration (i. e. 5 wt.% GNPs and 48 wt.% CB) contribute to the increase of the CB agglomeration resulting in a decrease of both mechanical strength and thermal conductivity (Figures 5b and 6b).

**Conclusions**

In this paper we adopt a processing technology to develop elastomer plus nano-graphite hybrid composites with multifunctional properties. Beyond the improvements of the mechanical properties, the research findings demonstrate the synergistic effect of carbon black and graphite nanoplatelets to prepare rubber composites thermally conductive and to design a new class of shock absorbers. It was found that a critical GNPs/CB ratio was able to reduce the strong interlayer forces among the GNPs sheets, which led to the efficiency on reinforcement in mechanical properties and improvements of the performance of the rubber composites.

**Acknowledgments**

N.M.P. is supported by the European Research Council (ERC StG Ideas 2011 BIHSNAM n. 279985 on Bio-Inspired hierarchical supernanomaterials, ERC PoC 2013-1 REPLICA2 n. 619448 on Large-area replication of biological antiadhesive nanosurfaces, ERC PoC 2013-2 KNOTOUGH n. 632277 on Supertough knotted fibers), by the European Commission under the Graphene Flagship (WP10 "Nanocomposites", n. 604391) and by the Provincia Autonoma di Trento ("Graphene Nanocomposites" , n. S116/2012-242637 and delib. reg. n. 2266).

Table 1. Recipes of the rubber compounds (indicated in phr: parts per hundred of rubber). The %weight content of GNPs/CB is reported below the name of each sample.

| Ingredient | EPDM-1 (0/0) | EPDM-2 (2/0) | EPDM-3 (5/0) | EPDM-4 (10/0) | EPDM-5 (0/48) | EPDM-6 (2/24) | EPDM-7 (5/48) |
|---|---|---|---|---|---|---|---|
| EPDM Vistalon 7500 | 100 | 100 | 100 | 100 | 100 | 100 | 100 |
| Paraffinic oil | 80 | 80 | 80 | 80 | 80 | 80 | 80 |
| Zinc oxide | 5 | 5 | 5 | 5 | 5 | 5 | 5 |
| Stearic acid | 1 | 1 | 1 | 1 | 1 | 1 | 1 |
| TMTD | 1.5 | 1.5 | 1.5 | 1.5 | 1.5 | 1.5 | 1.5 |
| Sulphur | 3 | 3 | 3 | 3 | 3 | 3 | 3 |
| Carbon black | --- | --- | --- | --- | 190 | 95 | 190 |
| Graphite Nanoplatelets | --- | 10 | 20 | 50 | --- | 10 | 20 |

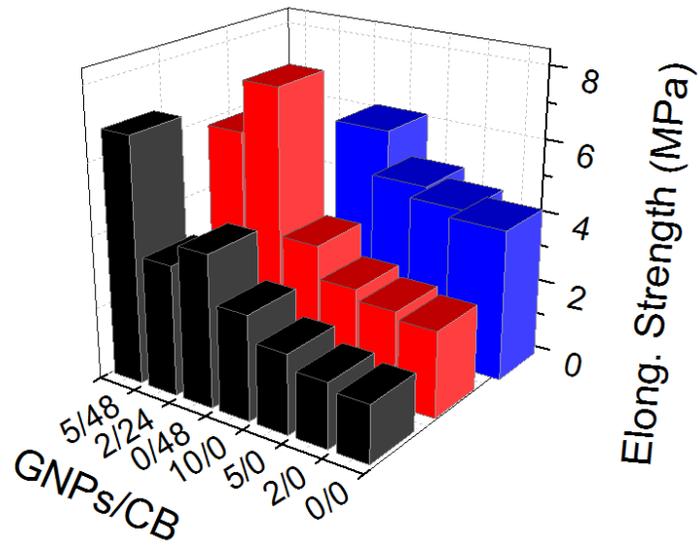

Figure 1. Modulus at different strains (black 50%, red 100% and blue 300%) for the prepared samples.

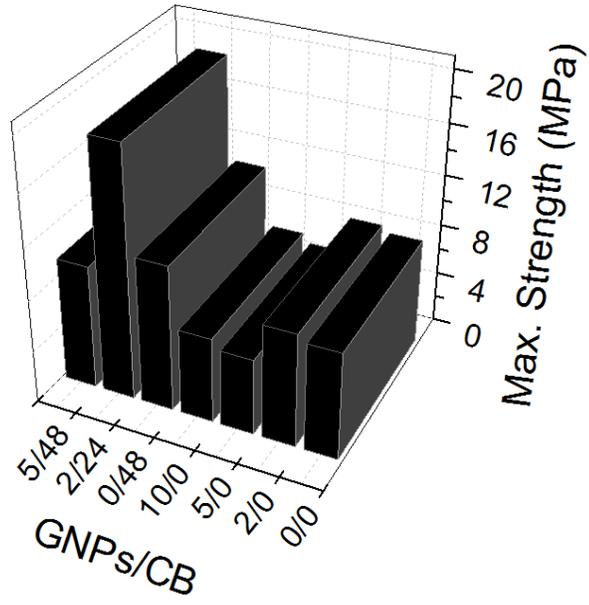

Figure 2. Maximum strength for the prepared samples.

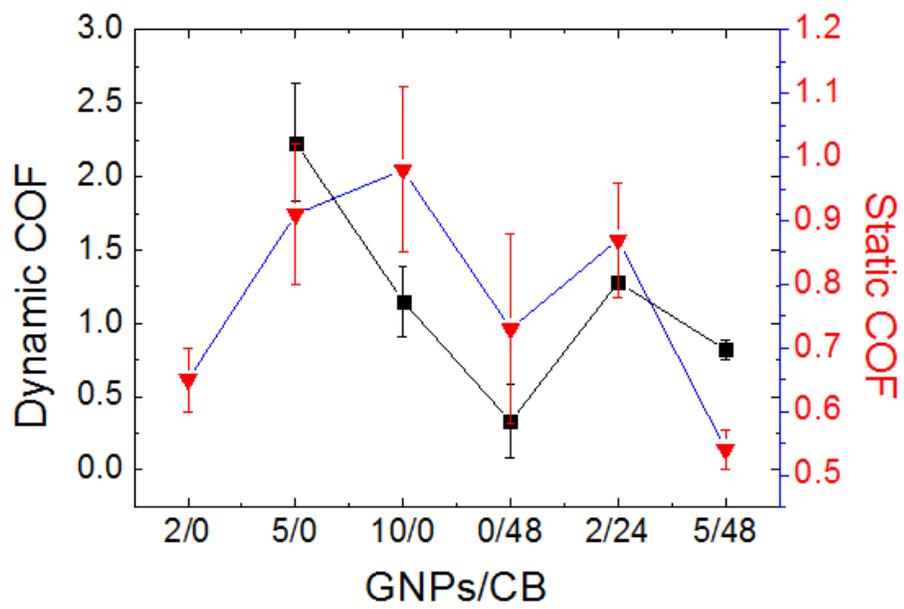

Figure 3. Dynamic and static coefficient of friction measured for the prepared samples.

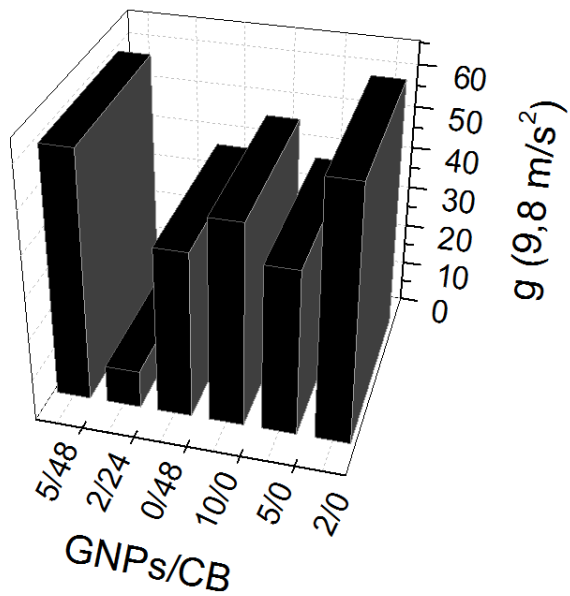

Figure 4. Accelerations transferred to the prepared samples by hitting them with a percussion energy of about 58J.

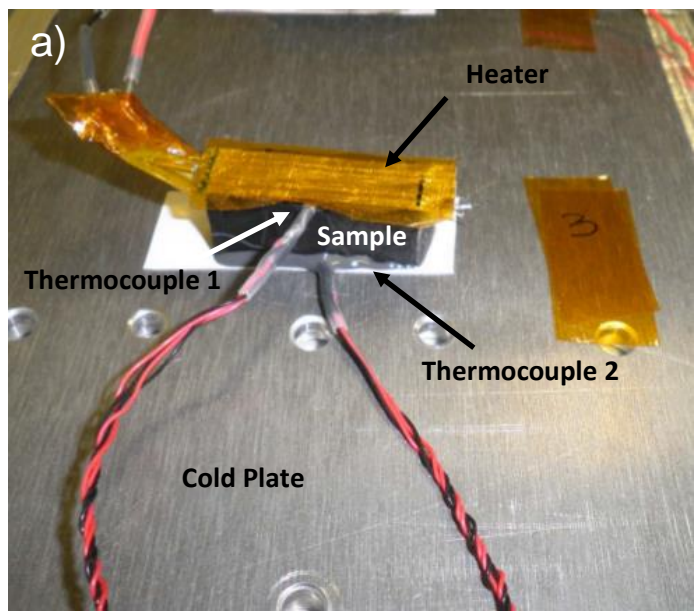

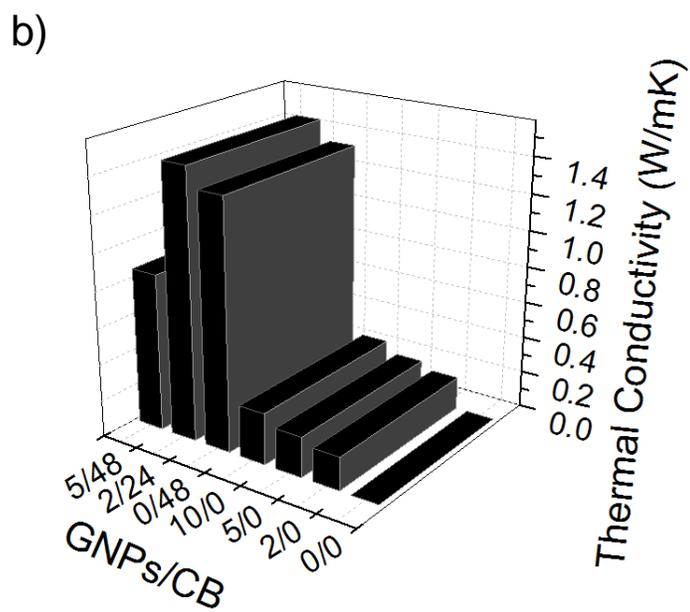

Figure 5. (a) Set up of the thermal conductivity measurements. (b) Thermal conductivity values as a function of the GNPs/CB content.

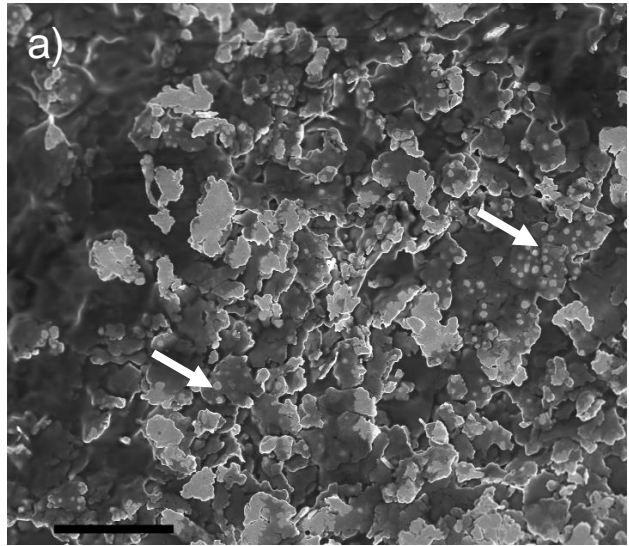

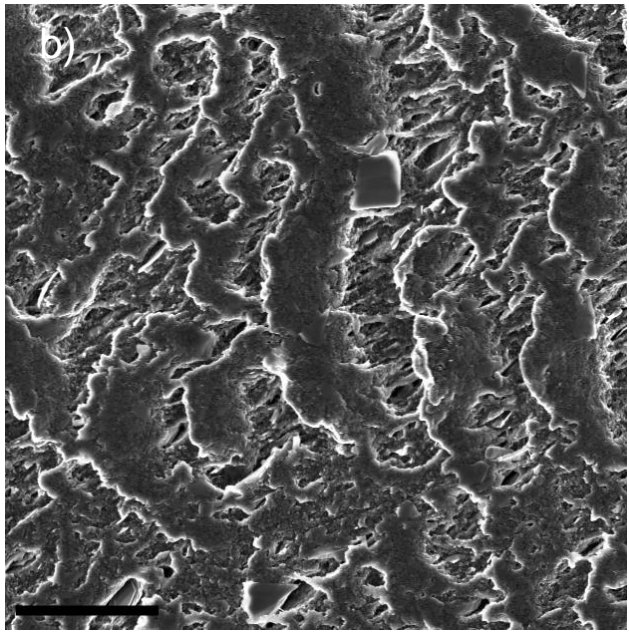

Figure 6. FESEM images of the a) EPDM-6 and b) EPDM-7 samples. The arrows on Figure 5a) shows the CB agglomerates on a GNP sheet. The scale bars indicate 1 μm.